**Exploring Fundamental Particle Acceleration and Loss Processes in Heliophysics through an Orbiting X-ray Instrument in the Jovian System**

*A White Paper for the 2024-2033 Solar and Space Physics (Heliophysics) Decadal Survey*

**Authors:** W. Dunn*[1,2], G. Berland*[3], E. Roussos[4], G. Clark[5], P. Kollmann[5], D. Turner[5], C. Feldman[6], T. Stallard[6], G. Branduardi-Raymont[24,2], E. E. Woodfield[7], I. J. Rae[8], L. C. Ray[9], J. A. Carter[6], S. T. Lindsay[6], Z. Yao[10], R. Marshall[3], A. N. Jaynes[11] A., Y. Ezoe[12], M. Numazawa[12], G. B. Hospodarsky[11], X. Wu[13], D. M. Weigt[14], C.M. Jackman[14], K. Mori[15], Q. Nénon[16], R. T Desai[17,18,7], L. W. Blum[3], T. A. Nordheim[19], J.U. Ness[20], D. Bodewits[21], T. Kimura[22], W. Li[23], H. T. Smith[5], D. Millas[1,2], A. D. Wibisono[24,2] , N. Achilleos[1,2], D. Koutroumpa[25] , S. C. McEntee[14] , H. Collier[26,27], A. Bhardwaj[28], A. Martindale[6], S.J. Wolk[29] S.V. Badman[9], R. P. Kraft[29]

1.University College London (UCL), UK 2. CPS UCL/Birkbeck, UK, 3. University of Colorado Boulder, 4. Max Planck Institute for Solar System Research, Germany, 5. Johns Hopkins APL, Laurel, MD, 6. University of Leicester, UK, 7. British Antarctic Survey, Cambridge, UK 8. Northumbria University, UK, 9. Lancaster University, UK, 10. Chinese Academy of Sciences, 11. University of Iowa, 12 Tokyo Metropolitan University, 13 University of Geneva, 14. Dublin Institute for Advanced Studies, Ireland, 15. Columbia University, NY, USA, 16. IRAP-CNRS, France, 17. Imperial College London, UK, 18. University of Warwick, UK, 19. JPL, California Institute of Technology, CA, 20. European Space Agency (ESA), ESAC, Spain 21. Auburn University, AL. 22. Tokyo University of Science, Japan. 23. Boston University, MA. 24. MSSL, UCL, UK. 25. LATMOS, CNRS, France. 26. ETH Zürich, Switzerland. 27. University of Applied Sciences and Arts (FHNW), Switzerland, 28. Physical Research Laboratory, India. 29 Smithsonian Astrophysical Observatory, CfA | Harvard & Smithsonian, MA. *These authors contributed equally to this work.

**Synopsis:** Jupiter's magnetosphere is considered to be the most powerful particle accelerator in the Solar System, accelerating electrons from eV to >70 MeV and ions to GeV energies. How electromagnetic processes drive energy and particle flows, producing and removing energetic particles, is at the heart of Heliophysics. Particularly, the 2013 Decadal Strategy for Solar and Space Physics was to *"Discover and characterize fundamental processes that occur both within the heliosphere and throughout the universe"*. The Jovian system offers an ideal natural laboratory to investigate all of the universal processes highlighted in the previous Decadal.

The X-ray waveband has been widely used to remotely study plasma across astrophysical systems. The *Chandra* and *XMM-Newton X-ray Observatories* alone have produced >14,000 refereed publications (>120,000 citations) to date. However, for <1% of these works the systems explored are within reach of in-situ measurements. The majority of astrophysical emissions can be grouped into 5 X-ray processes: fluorescence, thermal/coronal, scattering, charge exchange and particle acceleration. The Jovian system offers perhaps the only system that presents a rich catalog of all of these X-ray emission processes and can also be visited in-situ, affording the special possibility to directly link fundamental plasma processes with their resulting X-ray signatures. This offers invaluable ground-truths for astrophysical objects beyond the reach of in-situ exploration (e.g., brown dwarfs, magnetars or galaxy clusters that map the cosmos).

Here, we show how coupling in-situ measurements with in-orbit X-ray observations of Jupiter's radiation belts, Galilean satellites, Io Torus, and atmosphere addresses fundamental heliophysics questions with wide-reaching impact across helio- and astrophysics, including:

1. **What is the nature of the global structure and time-variability of the most intense radiation belts in the solar system, and what processes govern this?**
2. **How do losses of material to the atmosphere, moons, ring materials and neutral tori balance and ultimately limit radiation belt intensities?**
3. **How do X-ray signatures characterize astrophysical plasmas of different species?**
4. **Do rapidly-rotating giant planets, such as Jupiter, have a magnetospheric cusp and what is the extent and time-variability of this structure?**

From Earth's orbit, Jovian magnetosphere observations are limited by X-ray photon fluxes and spatial resolution. Soft X-ray (0.2-10 keV) imaging instruments have never flown to the realms of the gas giants. New developments like miniaturized X-ray optics and radiation-tolerant detectors, provide compact, lightweight (~10 kg), wide-field X-ray instruments perfectly suited to the Jupiter system, enabling this exciting new possibility. There is a fast growing heritage of such instruments on current/upcoming *NASA/ESA/JAXA/CAS* missions (*AXIS* on *AEPEX*; *MIXS* on *BepiColombo*; the *SXI* on *SMILE*; *MXT* on *SVOM*; *WXT* on *Einstein Probe*; *LEXT* on *Gamow)*. Here, we argue that such instrumentation is necessary to address fundamental heliophysics questions on radiation belt physics and relativistic particle behavior, thus enabling heliophysics to provide a unique and irreplaceable stepping-stone to astrophysical systems beyond the Solar System.

**Introduction:** In companion white papers submitted to the Decadal (Clark+, 2022; Kollmann+, 2022; Turner+, 2022), we present the case for exploration of the fundamental electromagnetic processes that govern particle acceleration and loss in heliophysics, by exploring the natural laboratories presented within the Jovian radiation belts, in particular through the NASA-funded mission concept *Comprehensive Observations of Magnetospheric Particle Acceleration, Sources, and Sinks* (*COMPASS* - Clark+, 2022). Here, we focus specifically on the use of X-rays to provide global context to single-point in-situ particle and field measurements, to offer new insights into the dynamics that govern such systems and to address fundamental questions in heliophysics.

Hitherto, all (0.2 - 20 keV) X-ray observations of the Jovian system were acquired by observatories in Earth-orbit. Since 1999, these were most commonly conducted by *NASA's Chandra* and *NuSTAR* X-ray observatories, *ESA's XMM-Newton* X-ray observatory and *JAXA's Suzaku* satellite (see Dunn, 2022 for review). These flagship X-ray observatories have provided revolutionary insights into the processes governing rapidly rotating magnetospheres, revealing: a variety of heavy ion aurorae (e.g., Branduardi-Raymont[B-R]+, 2008); magnetosphere - ionosphere (M-I) coupling through auroral hot spots (Gladstone+, 2002, *Nature*); clockwork-like electromagnetic pulsations (Dunn+, 2017, *Nature Ast.*); Jupiter's magnetar-like aurora (Mori+, 2022, *Nature Ast.*); X-ray fluorescence from plasma losses to moons, rings and tori (Bhardwaj+, 2005; Elsner+ 2002); radiation belt ultrarelativistic electron emissions (Ezoe+, 2010) and the potential universality of Electromagnetic Ion Cyclotron (EMIC) waves as a driver of ion aurorae (Yao+, 2021, *Science Adv.*). These breakthroughs were achieved despite limited Jovian X-ray fluxes at Earth-orbit assets, low spatial resolution and no access to Jupiter's nightside.

While Jupiter's magnetosphere glows in many wavelengths (visible, radio/synchrotron, UV, IR, visible and X-rays), what makes X-rays the standout waveband for radiation belt science is not only that their generation involves energetic particle interactions, but also that it is possible to distinguish the interactions of different particle species in different regions (moons, aurora, radiation belts, Io torus, jovian atmosphere), and map these to different energy ranges (from keV to relativistic energies). For example, where electrons can produce 0.2-20 keV (non-)thermal bremsstrahlung continuum emissions, colliding ion species each produce distinctive spectral lines (e.g., $O^{6+}$ ions are characterized by a triplet line at ~0.56 keV - Fig. 4b). In contrast, observations of e.g. Jupiter's synchrotron emissions only measure the low-altitude electron belt, offering little information on other species/regions. The X-ray spectrum depends both on the properties of the energetic particles (species and flux) and the scattering medium or neutrals. The X-ray spectra therefore provide unique insights into the energetic particle distributions (section 1 & 3), rates of collision and nature of the collisional material (section 2).

X-rays are thus the ideal waveband to globally observe the multiple components and processes of Jupiter's radiation belts. The archive of observations demonstrate that it is not only a feasible task, but that there is enormous potential for a step-change in understanding and for ground-breaking discoveries, through a dedicated Jupiter-orbiting X-ray imager (section 4).

**1. Identifying Global Dynamics & Source Processes through Time-Varying X-ray Spectra & Imaging of Jupiter's Radiation Belts:** The radiation belts are magnetospheric regions possessing high intensities of trapped energetic particles. Earth, Jupiter, Saturn, Uranus, Neptune, and Ganymede all possess radiation belts, suggesting that these may be a universal property of bodies with robust magnetospheres. Of the Solar System radiation belts, Jupiter's are the most intense, with >2 orders of magnitude higher intensities of >10 MeV electrons than the other planets (Mauk & Fox, 2010) and >GeV ions (Becker+, 2021). For Jupiter, the belts begin within 0.5 $R_J$ ($R_J$=Jupiter radii) of the cloud tops and continue beyond the orbit of Europa (at 9.4 $R_J$), dwarfing the scale and intensity of the entire closed terrestrial magnetosphere. In such a vast system, single-point in-situ particle/field measurements are key to identify driver processes, but such measurements require global context to identify which processes govern the system. Here, global images contemporaneous with in-situ measurements are essential.

The high energy regime of the Jovian system makes it an ideal target for X-ray observations. The 3 *Suzaku X-ray Imaging Spectrometer (XIS)* observations of Jupiter taken from 2006 to 2014 each independently revealed diffuse 1-5 keV X-ray emission from Jupiter's radiation belts (Fig. 1; Ezoe+, 2010; Numazawa+, 2019; 2021). The radiation belt X-ray spectrum shows that these emissions trace ultrarelativistic (>10 MeV) electron distributions - producing X-rays by Inverse Compton scattering of visible solar photons.

In tandem with in-situ measurements, this global imaging provides a powerful tool to test the sources for Jupiter's radiation belts. For a mission with approach-phase observations or a 30-90 $R_J$ apojove, this offers quasi-continuous coverage of the radiation belts, correlating currently unconstrained relationships between the belts and e.g. measurements of solar wind compressions, Io's volcanic activity or tail reconnection, addressing long-standing questions of the sources and processes regulating Jovian energetic particles. Comparison with global views of Earth's belts (from e.g. Van Allen Probes/SAMPEX) will reveal which processes are universal.

Acquiring images and spectra on sufficiently short timescales to connect these processes requires a Jupiter-orbiting X-ray instrument. Low X-ray fluxes arriving at Earth-orbit meant that each *Suzaku* detection required a 40-hr exposure. **At Jupiter orbit, Jovian X-ray fluxes are ~$10^7$ larger than fluxes reaching Earth**, enabling short-timescale (~hr) exposures that can measure time-variability in longitude, local-time and L-shell of the plasma distributions.

As is the case for other solar system X-ray sources (Lisse+, 2017; Dunn+, 2021), Jupiter's radiation belts are brighter than models predict (Numazawa+, 2021). Addressing these X-ray brightness inconsistencies is also key for interpretation of X-ray signatures across astrophysics. Solving this for Jupiter's radiation belts requires suitable, in-situ relativistic particle distribution measurements (Kollmann+, 2022) in tandem with X-ray observations, as proposed with *COMPASS* (Clark+, 2022). Such measurements will enable needed Inverse Compton models applicable to bodies from pulsars (Lyutikov+, 2012) to galaxies (Strong+, 2007 -1262 citations).

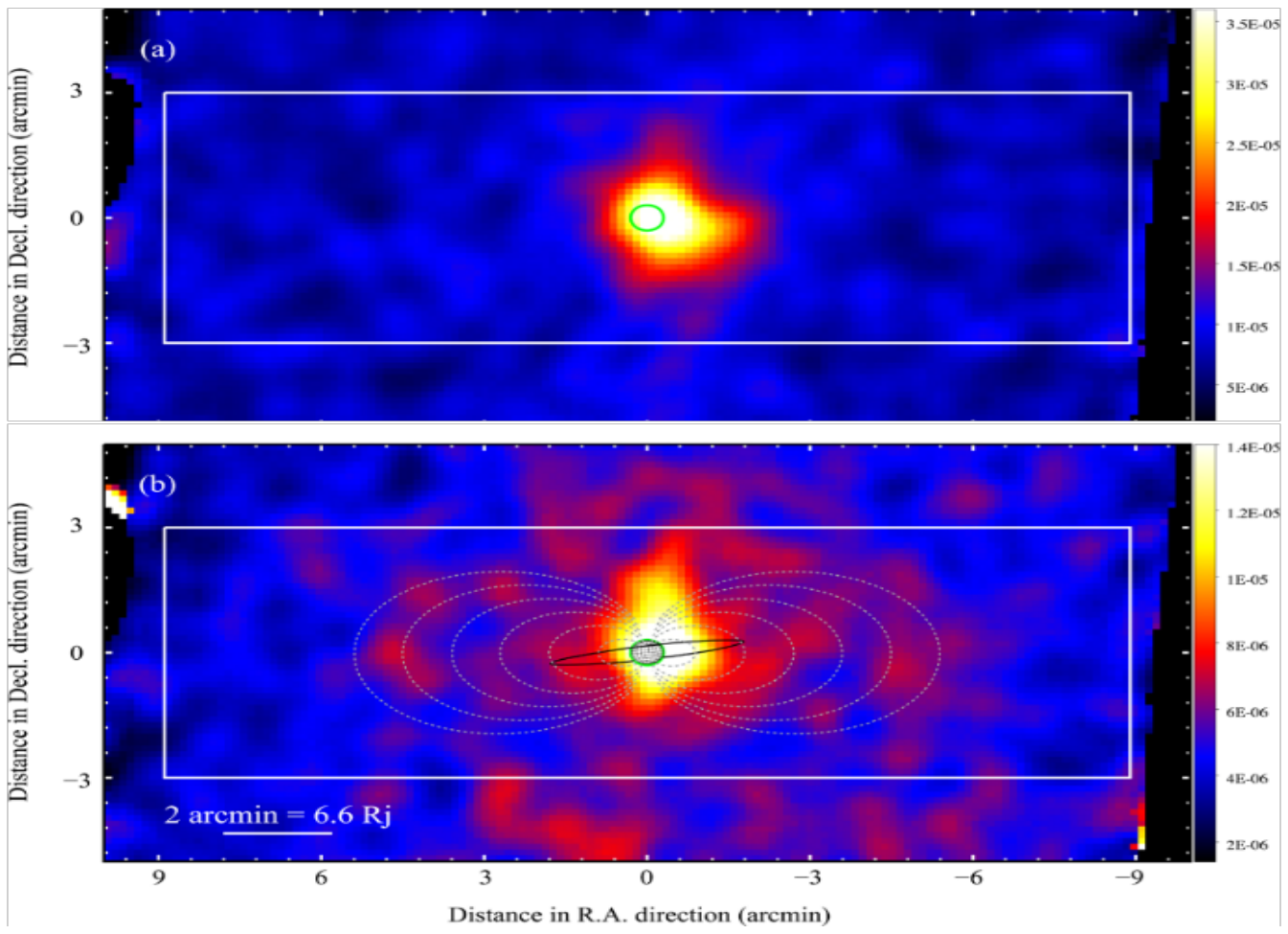

***Fig. 1: Suzaku XIS images of Jupiter and its Radiation Belts** (Numazawa+, 2019): in the energy range 0.2–1 keV (a) and 1–5 keV (b), showing electron emissions from Jupiter's radiation belts at 1-5 keV. The green circle indicates Jupiter's position and size (39") at the time. Grey dotted lines show magnetic field lines with equatorial crossings of 3, 6, 9, 12, 15, and 18 $R_J$. The black line is Io's orbital path.*

**2. Characterizing Radiation Belt Losses through X-ray maps of Jupiter's Moons, Tori and Rings:** While acceleration and source processes are a key focus in radiation belt physics, losses are also important because without them intensities would accumulate indefinitely. For Jupiter, the radiation belts are embedded deep in the inner magnetosphere (<15 $R_J$), far from the magnetopause standoff distance (60-100 $R_J$). Consequently, unlike for Earth, Jupiter's radiation belt particles cannot directly escape to the solar wind (SW). Instead, processes within the inner magnetosphere control the state and losses of the belt population. Across the solar system, moons, neutrals and rings modulate, erode and destroy radiation belts (Kollmann+, 2017; Roussos+, 2018). Jupiter's inner magnetosphere is rich with moons, rings and gas tori that not only supply plasma, but also remove energetic particles from the system (Fig. 2). Why the balance of processes favors radiation build-up is not understood. X-ray emissions provide the key: remotely monitoring and diagnosing energetic particle collisions with neutrals, to identify, map and characterize global loss processes and their time-variability.

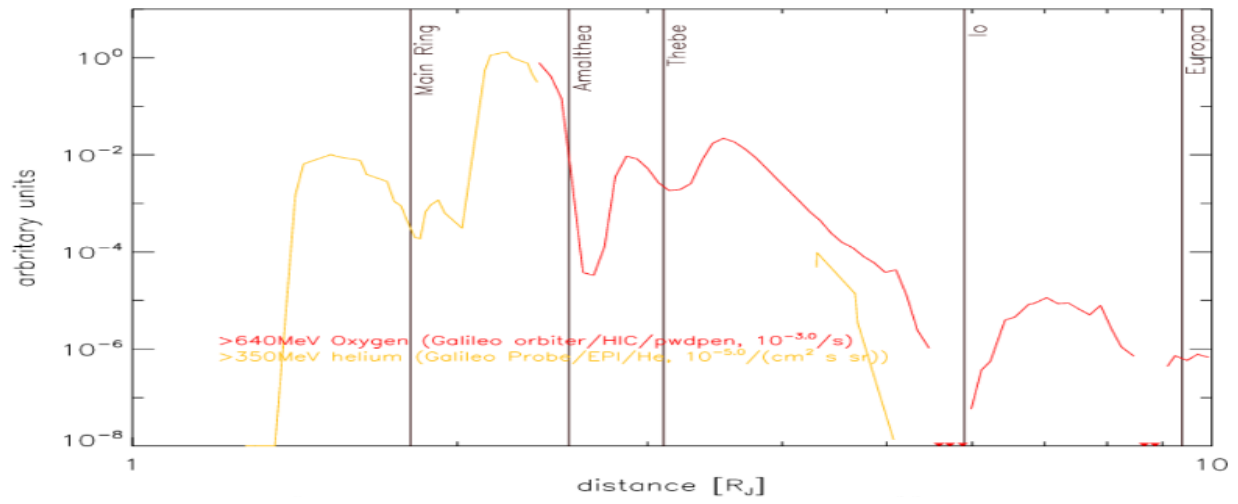

***Fig. 2: The removal of Jupiter's radiation belts by moons and the main ring*** *(Kollmann+, 2022): Energetic Oxygen and Helium intensities throughout the Jovian radiation belts, showing the reduction in energetic particles at the orbits of the moons and ring.*

In 1999, the first *Chandra* X-ray observations of Jupiter revealed the presence of X-ray emissions from Io (Fig. 3), Europa, and the Io Torus (Elsner+, 2002). X-ray emissions from the Galilean satellites are produced when energetic particles collide with the moons' surfaces (Nulsen+, 2020). The resulting X-ray emissions are therefore dependent on the composition of the surface and the contributions of bombardment by ions and electrons (Paranicas+, 2002). Ion collisions produce Particle Induced X-ray Emission (PIXE – Johansson+, 1995), characterized by strong, discrete fluorescence lines. In contrast, electrons produce thick-target bremsstrahlung (TTB), which has continuum emissions, and can also produce PIXE lines (mostly for Z > 13 elements) (Markowicz & Van Grieken, 1984). Precipitation of particles onto the moons' surfaces depends on their pitch angle and thus on the plasma population, magnetic field configuration and wave-particle interactions (Nénon+, 2017; 2018) so that X-ray emissions track the magnetospheric drivers of precipitation.

From Earth orbit, detectable X-ray fluxes from the Galilean moons are very low, so that it is impossible to explore variation with time/hemisphere. While studies from Earth-orbit assets have reached their ceiling, the ~$10^7$ X-ray flux increase available to a Jupiter-orbiting instrument transforms capabilities for the Galilean moons, enabling time variability and maps of energetic particle losses to be studied simultaneously with in-situ measurements of the belt conditions.

Unlike Saturn, where X-ray fluorescence from the rings is observed (Bhardwaj+, 2005), X-ray emissions from radiation belt collisions with Jupiter's more tenuous rings, and moons Amalthea and Thebe (Fig. 2), are yet to be explored from Earth-orbit, due to signal limitations. X-ray emissions from the Io Torus have also been observed from Earth, but their cause is unclear, seemingly due to a combination of electron- and ion-neutral collisions (Elsner+, 2002) that track the role of tori in loss processes. The low-signal at Earth-orbit makes interpretation of Io torus variation challenging. This is again solved through orbiting X-ray instrumentation.

Across the wider domain of astrophysics, X-rays are used to study dusty systems (e.g., inter/stellar objects and nebulae - see, Heinz & Corrales, 2016 for review), through the interaction of energetic particles with neutrals. For such systems, the observations proposed here provide critical ground-truths to enable interpretation of solely remote signatures. However, no Earth-orbiting X-ray instrument will provide improved spatial resolution for the foreseeable future. E.g., *ESA's* next-generation X-ray observatory, *ATHENA+* (launch 2030s), has an order of magnitude lower spatial resolution than *Chandra*. Breakthroughs from mapping and characterizing particle losses to the moons require a Jupiter-orbiting X-ray instrument.

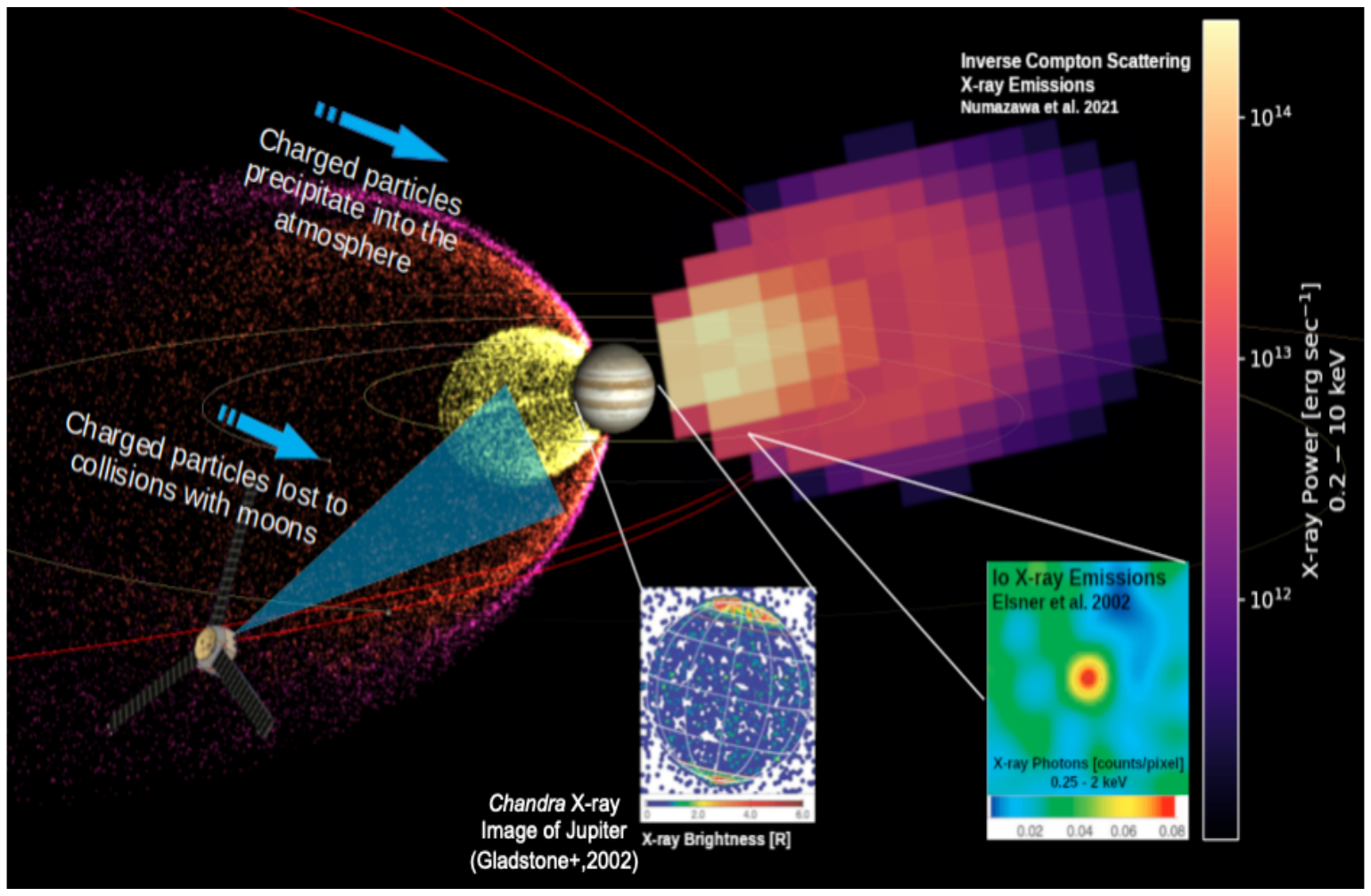

***Fig 3.*** *Schematic of observations by the COMPASS X-ray Imager of: charged particle precipitation into the Jovian atmosphere (with insert Chandra observation showing Jupiter's auroral and equatorial X-ray emissions - Gladstone+, 2002); charged particle losses to the moons (with insert Chandra observation of Io from Elsner+, 2002) and modeled Inverse Compton X-rays from Jupiter's radiation belts.*

**3. Atmospheric Belt Losses, M-I Coupling & the Universality of Open Magnetospheric Systems:** Energetic particles precipitating into Jupiter's atmosphere produce an array of X-ray emissions. Depending on the precipitating particle, different X-ray spectral signatures are produced. Electron precipitation produces bremsstrahlung, diagnostic of the electron distribution (B-R+, 2004; Mori+, 2022, *Nature Ast.*), while sufficiently charged or >300 keV/amu ions produce CX lines (Fig. 4), identifying the precipitating ion species and energies (B-R+ 2007; Houston+, 2020).

Like Earth, Jupiter loses radiation belt particles via precipitation to the atmosphere (Kollmann+, 2021). While Jupiter's low-latitude X-ray emissions are dominated by solar photons that elastically scatter off atmospheric hydrogen (Maurellis+, 2000), there are also emission enhancements where the radiation belt loss cone is largest (Bhardwaj+ 2006; McEntee+, in review). From Earth-orbit, where Jupiter's sunlit face is the only observable hemisphere, the X-ray spectrum from this radiation belt precipitation cannot be distinguished against the dominant scattered solar photon signal. However, for Jupiter-orbiting spacecraft, nightside observations - free from scattered sunlight - can monitor spectral/spatial/temporal characteristics of electron and ion belt losses comparable with simultaneous measurements of e.g. EMIC/whistler waves (Nénon+, 2017; 2018); connecting source and loss processes.

Jupiter possesses the most powerful aurora in the Solar System, which dominates its high latitude X-ray emissions (Fig. 4). As with pulsars, Jupiter's aurorae are tilted to the rotation axis and fixed in the rotating frame so that they rotate in and out of view with planetary rotation. While many wavebands provide key insights into Jupiter's aurorae, the X-rays stand out by showing emission from the precipitating particles and not the atmospheric response (e.g. UV/IR aurora) and by distinguishing electron aurora (Bremsstrahlung) from heavy ion aurora (CX lines).

This offers a holistic characterization of M-I coupling, tracking upward and downward currents. Particularly, Jupiter's heavy ion aurorae are a unique Solar System laboratory, providing insights into M-I coupling in magnetospheres with high heavy ion abundances. Observations hint at arcs, time-variable pulsations and boundaries, that change on timescales too rapid for low-signal (~1 count/min) Earth-orbit observations to resolve (Dunn+ 2016). The proposed *COMPASS X-ray Imager (XRI)* would observe a truly transformative planetary signal of ~10,000 counts/second, revealing the morphology of what are currently sparse clusters of dots.

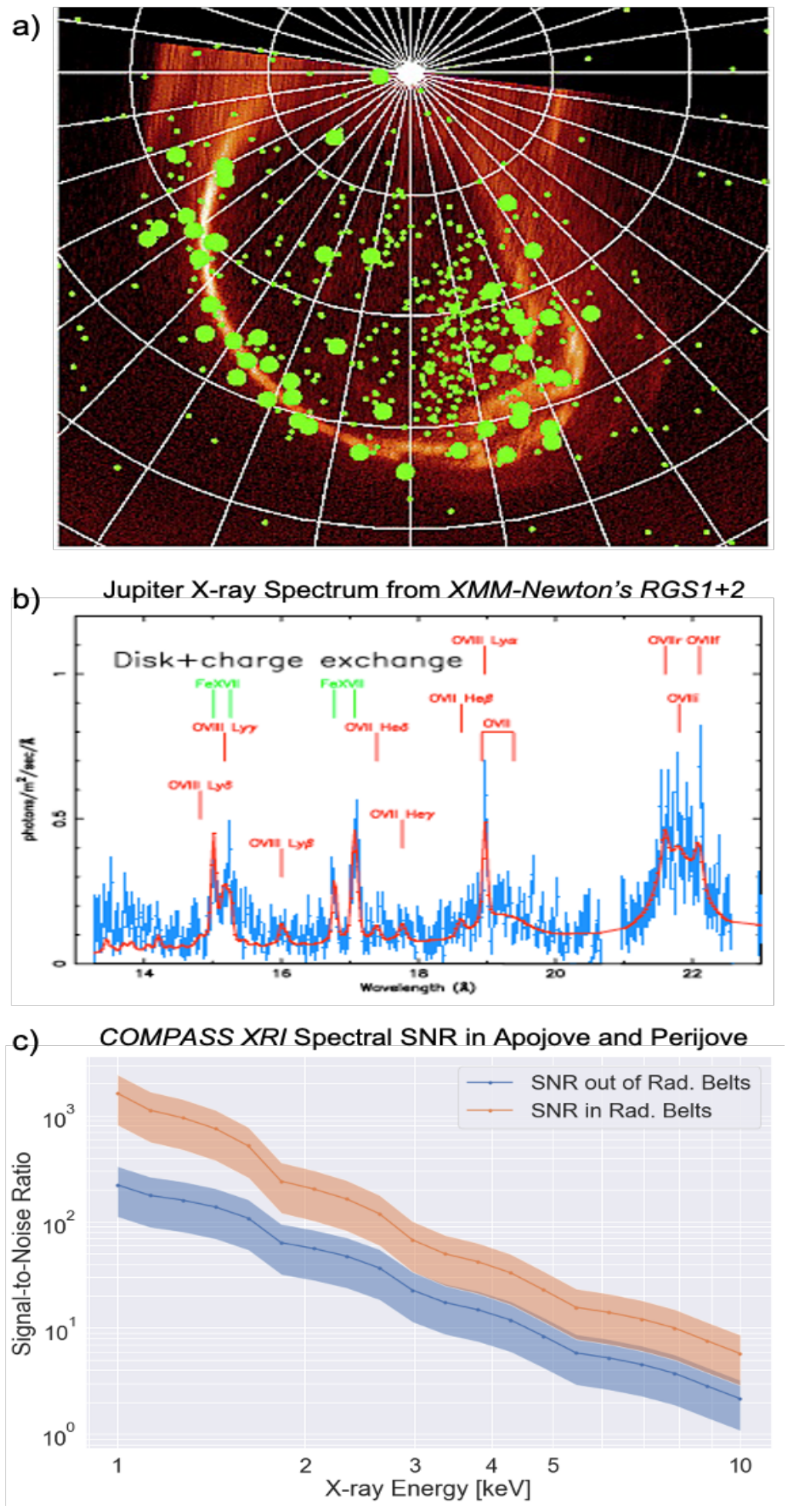

***Fig. 4:*** *a) Projection of Jupiter's North pole showing the UV (orange) and X-ray (green dots) aurorae as recorded by the Hubble Space Telescope STIS FUV instrument and Chandra ACIS instrument, on 24 Feb 2003. Each green dot indicates a single X-ray photon. Small green dots show <2 keV photons - predominantly CX lines from precipitating heavy ions. Large green dots show >2 keV photons from electron bremsstrahlung. Bremsstrahlung is co-located with the UV main oval, while the heavy ion aurorae occur poleward of this. The 10° latitude-longitude spaced grid is fixed in Jovian System III coordinates with 180° toward the bottom and 270° to the left (B-R+, 2008). b) XMM-Newton high resolution spectra from Jupiter (blue crosses), combining the RGS 1+2 spectra. Overlaid in red is the best fit model, which includes a CX model, representative of Jupiter's soft X-ray aurorae, and scattered solar X-ray lines to represent the Jovian disk emission (B-R+, 2007) c) Jupiter X-ray aurora Signal to Noise Ratio in and out of the radiation belts with the COMPASS X-ray Imager (XRI).*

In heliophysics, it is critical to understand to what extent magnetospheres are universally magnetically open or closed, and the importance of the Dungey cycle for different systems. For rapid-rotating systems like Jupiter, the timescale over which open magnetic field lines convect into the tail is much longer than the planet's rotation, leading to discourse over whether a Dungey cycle is feasible at all for such systems (McComas & Bagenal, 2007; Cowley+, 2008). With heliospheric distance, systems may transition from Dungey-dominated (e.g. Mercury) to viscous-dominated SW-interactions (e.g. Jupiter) (Delamere & Bagenal 2010; Masters, 2018).

X-ray observations may uniquely solve the longstanding mystery of the extent to which Jupiter is open to the SW. Through SWCX emissions, in which SW ions collide with neutrals and produce CX lines, it is possible to identify intervals when SW precipitates into the atmosphere. On rare occasions, X-ray observatories do detect SWCX signatures from Jupiter (Dunn+, 2020b). However, it has been impossible to identify their source region and duration due to the low spatial resolution and signal at Earth-orbit. This problem vanishes for the photon-rich signal available for a Jupiter-orbiting instrument. By utilizing SWCX signatures, the *ESA-CAS SMILE* spacecraft will study Earth's cusps in this way (Sibeck+,2018; B-R+,2018). Similar searches for SWCX from Jupiter's cusp will test the presence, orientation and time variability of Jupiter's cusp and the extent to which rapidly-rotating and/or distant magnetospheres are open to the SW.

**4. New Advances in X-ray Instrumentation that Enable Ground-Breaking Heliophysics Studies:** Recent step-changes in X-ray instrumentation such as Micro-Pore Optics and radiation tolerant detectors open the possibility of compact, lightweight (~10 kg), low power (~W), X-ray instruments perfectly-suited for Heliophysics science at Jupiter. To observe the sources discussed in this paper, an instrument requires: an energy range of 0.2-8 keV with resolution of 0.15 keV at 0.56 keV; Field of View (FoV) of 22+/-5° (global views of belts from 30-90$R_J$); angular resolution of 1° to access L-shell resolution on Jupiter during perijove[PJ] (4-6 $R_J$) and to resolve the moons at apojove, or ~10' resolution to resolve surface features on moons from <10 $R_J$.

Several instruments in flight or due for launch within 5 years carry lightweight X-ray instruments that meet the requirements outlined above and will achieve related science goals, including: *MIXS* on *BepiColombo,* to observe PIXE and TTB from Mercury (Bunce+, 2020); the *SXI* on the *SMILE* mission, to study SWCX from the terrestrial cusps and magnetosheath (B-R+, 2018) or the Chinese-French SVOM mission to observe Gamma Ray Bursts. *JAXA's* Jupiter-orbiting *JUXTA* X-ray instrument concept, also laid essential groundwork (Ezoe+, 2013).

For the *NASA* heliophysics mission-concept *COMPASS*, the onboard *X-ray Imager (XRI - mass including shielding: 15 kg; power: 15 W)* is based on *AEPEX/AXIS* heritage (Marshall+, 2020). The XRI provides a 22° FoV with 1° resolution (50-200 km moon resolution during flybys; L-shell resolution on Jupiter during PJ). While modern radiation-tolerant X-ray detectors can be built to survive Jupiter's particle background, noise is a challenge. Fig. 4c shows that despite this the signal:noise will achieve the science goals. To handle the extraordinary PJ X-ray signal of 10,000 cts/s, the instrument connects 16 X-ray detectors (each handling $6 \times 10^4$ cts/s).

Jupiter-orbit X-ray instrumentation and science is now high feasibility and low-risk, opening an exciting new paradigm in heliophysics. An unexplored treasure trove of discoveries awaits Jupiter-orbit X-ray instrumentation, which is essential for addressing an array of fundamental questions on the transfer of energy and particles in high energy systems. Through such observations, heliophysics will provide an irreplaceable stepping-stone to astrophysical systems beyond our solar system.